\documentstyle[multicol,aps, prl]{revtex}

\date{\today}
\draft
\tighten
\begin{document}
\def\sqr#1#2{{\vcenter{\hrule height.3pt
      \hbox{\vrule width.3pt height#2pt  \kern#1pt
         \vrule width.3pt}  \hrule height.3pt}}}
\def\square{\mathchoice{\sqr67\,}{\sqr67\,}\sqr{3}{3.5}\sqr{3}{3.5}}
\def\today{\ifcase\month\or
  January\or February\or March\or April\or May\or June\or July\or
  August\or September\or October\or November\or December\fi
  \space\number\day, \number\year}

\def\Bbb{\bf}
\topmargin=-0.1in


\newcommand{\ww}{\mbox{\tiny $\wedge$}}
\newcommand{\pp}{\partial}

\title{Thermal Equilibration of Brane-Worlds}

\author{Andrew Chamblin, Andreas Karch, and Ali Nayeri}

\address {\qquad \\ Center for Theoretical Physics \&
Department of Physics\\
Massachusetts Institute of Technology\\
77 Massachusetts Avene\\
Cambridge, MA 02139, USA
}

\maketitle

\begin{abstract}

We analyze the thermodynamical properties of brane-worlds, with
a focus on the second model of Randall and Sundrum.  We point out
that during an inflationary phase on the brane, black holes will tend
to be thermally nucleated in the bulk.  This leads us to ask
the question:  Can the black hole - brane-world system evolve towards 
a configuration of thermal equilibrium?  To answer this,
we generalize the second Randall-Sundrum scenario to allow for
non-static bulk regions on each side of the brane-world.  Explicitly,
we take the bulk to be a {\it Vaidya-AdS} metric, which
describes the gravitational collapse of a spherically symmetric null
dust fluid in Anti-de Sitter spacetime.  Using the background subtraction
technique to calculate the Euclidean action, we argue that 
at late times a sufficiently large black hole will relax to a point of
thermal equilibrium with the brane-world environment.  
These results have interesting implications for early-universe cosmology.

\end{abstract}


\begin{multicols} {2}

\section{Introduction: The Thermodynamics of inflating brane-worlds}

Recently \cite{RS2}, Randall and Sundrum presented a model
in which the universe is realized as a ${\bf Z}_{2}$-symmetric
positive tension domain wall, or brane, in $AdS_5$.  They analyzed the 
linearized equation for a graviton propagating in this spacetime.
They proved that there is a solution
describing a `bound state', i.e., an integrable wave function corresponding
to a massless graviton which is confined to the domain wall.
For low energy processes this bound state dominates
over the Kaluza-Klein (KK) states, so that 
Newtonian gravity is recovered as long as the length scale of the
AdS space is sufficiently small.

Soon after this model appeared,
it was pointed out \cite{gubser} that the AdS/CFT correspondence
gives rise to the Randall-Sundrum model in a certain limit.
More precisely, the AdS/CFT correspondence relates the
Randall-Sundrum model to an equivalent four dimensional theory
consisting of general relativity coupled to a strongly interacting
conformal field theory.

Now, the Hawking-Page phase transition
\cite{hawkpage} manifests itself in the AdS/CFT duality (\cite{ed}).  
Explicitly,  there is a critical temperature,
$T_c$, past which thermal radiation in AdS is unstable to the formation of a
Schwarzschild black hole.  (In fact, for $T{>}T_c$
there are two values of the black hole mass at which the Hawking
radiation can be in equilibrium with the thermal radiation of the
background.  The lesser of these two masses is a point of unstable
equilibrium (it has negative specific heat), whereas the greater mass
is a point of stable equilibrium.)

Since the second Randall-Sundrum model may be understood
by looking at the AdS/CFT
duality in the non-gravity decoupled limit, one would expect that a similar
phase transition should occur in that setting.  In particular, during an
inflationary phase on the brane, the brane-world is a
de Sitter hyperboloid embedded in $AdS_5$, and it will generate an
acceleration horizon in the bulk.  This horizon will have a temperature,
and so we would expect that inflating brane-worlds would be unstable
to the creation of bulk (five-dimensional) black holes.
In fact, this process was discussed
in a recent paper by Garriga and Sasaki \cite{garriga}.  There, the authors
studied the `thermal instantons' which correspond to black
holes in AdS, and showed that these instantons describe the
thermal nucleation of Schwarzschild-AdS black holes in the presence of 
a pre-existing ${\Bbb Z}_2$ symmetric inflating brane-world.

For us, this is evidence that we should {\it assume} that a bulk
black hole will be formed during the inflationary epoch of a brane-world universe.
It therefore behooves us to answer the question: Can an inflating brane-world and a
bulk black hole ever attain thermal equilibrium?

This question is very subtle, because we have to decide what boundary 
conditions to impose on the bulk fields at the location of the brane-world.
Here we assume that the brane-world
is a ${\Bbb Z}_2$ symmetric domain wall, so that the brane acts
like a reflecting mirror for massless bulk modes.  In other
words, for us the brane-world
is like an accelerating mirror: A `black box' with perfectly
reflecting, accelerating walls.  Certainly, this is the standard boundary
condition for bulk modes when the wall is located at the boundary of
AdS, and in particular it is the condition assumed by Hawking and Page
\cite{hawkpage}.  Because the wall acceleration will not be constant
(as the wall climbs out of the potential well of the black hole),
the wall should emit a flux of radiation with some temperature.
There are thus naively three basic cases to consider.  First, the black
hole temperature ($T_{BH}$) may be greater than the brane-world temperature
($T_{BW}$), in which case the hole has negative specific heat and it will
evaporate in a finite time and have no effect on the late time evolution of 
the system.  Alternatively, the system may be fine-tuned so that 
$T_{BH} = T_{BW}$ exactly; this would describe thermal equilibrium between
the two competing temperatures.  Finally, it may be the 
case that $T_{BW} > T_{BH}$\footnote{This
analysis is similar to that of Yi \cite{piljin}, who studied black holes
which are uniformly accelerated by cosmic strings.}; clearly, this
case is of interest because it would appear that the black hole
might be unstable to some runaway process where the mass increases
indefinitely.  One approach to this problem is to actually
explicitly compute the brane-world temperature exactly; however, this task
is rather difficult and one is inevitably led to estimates which are
hard to control.  In this paper, we will instead approach this problem
using the more elegant techniques of Euclidean quantum gravity.  In order
to motivate this analysis, we will begin with a discussion of brane-world
evolution in a non-static bulk with collapsing null dust fluid.
We work with the signature convention $(-,+,+,+,+)$.

\section{Gravitational collapse in anti-de Sitter space: The Vaidya metric}

In 1951, Vaidya \cite{vaidya} wrote down a metric that represents an
imploding (or exploding) null dust fluid with spherical symmetry in asymptotically
flat space.  Recently \cite{rio}, this metric has been generalized to 
describe gravitational collapse in spacetimes with a non-vanishing 
cosmological constant.  Here, we are interested in the metric which
describes gravitational collapse in a spacetime which is asymptotically
Anti-de Sitter (AdS).  This metric is written using 
`Eddington-Finkelstein' coordinates, so that it takes the explicit form
\begin{equation}
\label{vaidya}
ds^{2} =  - e^{2\psi(v, r)} f(v,r)dv^{2} +
2 \epsilon e^{\psi(v, r)}dvdr\\
 + r^{2}d\Omega^{2}_{3},
\end{equation}

\noindent where
\[
f(v,r) = k - \frac{2M(v, r)}{r^2},
\]

\noindent and 
\[
d\Omega^{2}_{3} = d\chi^2 + R_{k}^{2}(\chi)
\left(d\theta^2 + \sin^2{\theta} 
d\phi^2\right),
\]
with $R_{-1}(\chi) = \sinh{(\chi)}$, $R_{0}(\chi) = \chi$ and
$R_{+1}(\chi) = \sin{(\chi)}$
is the metric on hyperbolic space, flat space and the round three-sphere
respectively.
The function $M(v,r)$, called the {\it mass function},
is a measure of the total gravitational energy 
within a radius $r$.  The sign $\epsilon = \pm 1$ indicates whether the
null coordinate $v$ is advanced or retarded.  If $\epsilon = + 1$
then $v$ is advanced time, in which case rays
of constant $v$ are ingoing.  Likewise, $\epsilon = - 1$ means that
rays of constant $v$ are outgoing.  Since we are interested in collapsing
radiation, we will assume $\epsilon = +1$.

In \cite{rio} the authors take $\psi(v, r) = 0$, so that the Einstein
equations simplify considerably.  We also make this assumption, 
and so we may follow
their analysis and conclude that the source for this metric is a 
`Type II fluid' \cite{stephen}.
Following \cite{rio}, it is straightforward to see that if we want to 
describe gravitational collapse (or the time reverse)
in $AdS_5$ then the appropriate mass function, in general, is
\begin{equation}
M(v, r) = \frac{\Lambda}{12}r^{4} + m(v) - \frac{q^2(v)}{r^2},
\end{equation}

\noindent where ${\Lambda} = -(6/l^2)$ is the bulk cosmological
constant, and $m(v)$ is an arbitrary function of $v$ which will be determined
by the energy density of the radiation in the bulk.  $q(v)$ corresponds to
the charge of the bulk, if any.

Now that we have a precise idea of what the non-static bulk metric looks
like, we turn to the question of how a domain wall moves in such a bulk.

\section{The Israel Equations of Motion: The cosmology of brane-worlds}

The equations of motion for a domain wall, when the effects of 
gravity are included, are given by the Israel junction conditions.
These conditions relate the discontinuity in the extrinsic curvature 
($K_{A B}$)
at the wall to the energy momentum ($t_{A B}$) of fields which live
on the wall:
\begin{equation}
\label{eqn:israel}
\left [K_{A B} - Kh_{A B}\right]^{\pm} = \kappa^{2}_{D} t_{A B}.
\end{equation}
(see \cite{dddw} for a derivation of this equation).
The gravitational coupling constant, $\kappa^{2}_{D}$, in arbitrary dimension $D$,
is given by~\cite{rman}
\begin{equation}
\kappa^{2}_{D} = \frac{2(D - 2)\pi^{\frac{1}{2}(D - 1)}}
                      {(D - 3)(D/2 - 3/2)!}G_D, 
\end{equation}
where $G_D$ is the $D$-dimensional Newton constant.  Here,
for instance, $\kappa^{2}_{5} = 3 \pi^2 G_5$.
Given the form of (\ref{eqn:israel}), it is obvious that we need to 
calculate the extrinsic curvature for timelike hypersurfaces which 
move in the Vaidya-AdS background.

As above, we begin by writing the metric in Eddington-Finkelstein
coordinates:
\begin{equation}
\label{eqn:eddington}
ds^2 = dv \left[ -f(v,r)dv + 2dr \right]
+ r^{2}d\Omega^{2}_{3},
\end{equation}

\noindent where we have assumed that the null coordinate
$v$ represents Eddington advanced time.  Since we are interested in the
cosmological aspects of brane-world evolution, we want the metric
induced on the brane-world to assume a manifestly FRW form:
\begin{equation}
\label{eqn:frw}
ds^{2}|_{brane-world} = 
- d\tau^{2} + a^2(\tau)d\Omega^{2}_{3},  
\end{equation}
\noindent where the coordinates $x^{\mu} = ({\tau}, {\chi}, {\theta}, {\phi})$
are the coordinates intrinsic to the brane-world.
This means that we constrain the timelike hypersurface describing the evolution
so that the brane can only shrink or contract as it moves in time, i.e.,
the position of the brane at a given time should be completely specified
by its radial position: $r = r(v) = a(\tau)$.
(In what follows we will let $a = a(\tau)$ denote the position of the
brane, in order to avoid possible 
confusion between the coordinate $r$ and
the radial position of the brane-world).
In this way, the problem is basically reduced to a 1+1 dimensional
system.  Furthermore, we need only focus on the contributions to the 
Riemann tensor which are induced by the $(r,v)$-sector of the metric.
Using the fact that $\tau$ is the
time experienced by observers who move with the brane-world:
\begin{equation}
\label{eqn:comoving}
d\tau = \left(f - 2\frac{da(\tau)}{dv}\right)^{1/2}dv,
\end{equation}

\noindent we may express the problem in the $(\tau,a)-sector$.
In particular, if we let `$\dot{a}$' denote differentiation relative to 
comoving
time, i.e., $\dot{a} = da/d\tau$, then one can find the extrinsic
curvature through the following relation

\[
K_{\mu \nu} = - e^{A}_{\mu} e^{B}_{\nu} \nabla_A n_B,
\]
with
\[
e^{A}_{\mu} = \left(\frac{1}{f}\left(\dot{a} + \sqrt{\dot{a}^2 + f}\right)
 \delta^{A}_{v} + \dot{a} \delta^{A}_{r}\right)\delta_{\mu\tau}
 + \delta^{A}_{\mu},
\]
as the tetrads at the wall and
\[
n_A = - \dot{a} \delta^{v}_{A} + \frac{1}{f}\left(\dot{a} + \sqrt{\dot{a}^2 
+ f}\right) \delta^{r}_{A},
\]
as the unit normal vector to the hypesurface $a(\tau)$.  $\nabla_A$ is the
covariant derivative associated with the metric (\ref{eqn:eddington}).
The nonvanishing components of extrensic curvature are then

\begin{eqnarray}
\label{eqn:ttextrinsic}
K^{\tau}_{\tau} & = & - \frac{1}{2}\frac{2 {\ddot a} + f'}{\sqrt{f + 
{\dot a}^2}}
+ \frac{1}{2}\frac{\dot{f}}{f^2}\left(\dot{a} + \sqrt{f + \dot{a}^2}\right)^2
\nonumber \\
                & = & -\frac{d}{da}\left({\sqrt{f + 
{\dot a}^2}}\right)
+ \frac{1}{2}\frac{\dot{f}}{f^2}\left(\dot{a} + \sqrt{f + \dot{a}^2}\right)^2
\end{eqnarray}

\begin{eqnarray}
\label{eqn:angular}
K^{\chi}_{\chi} = K^{\theta}_{\theta} = K^{\phi}_{\phi} = 
- \sqrt{H^{2}(\tau) + \frac{f(\tau,a)}{a^2}}~,
\end{eqnarray}
with $H \equiv (\dot{a}/a)$.

Given these expressions, we can examine how the non-static
bulk is affecting cosmology.  That is, we assume that the stress-energy tensor 
describing the fields which propagate in the brane-world is given as
\[
t^{A}_{B} = \mathrm{diag}(-(\rho+\rho_{\lambda}),p-\rho_{\lambda},
p-\rho_{\lambda},p-\rho_{\lambda},0)
\]

\noindent We emphasize that $\rho$ and $p$ are the energy density and pressure
of the ordinary matter, respectively, whereas, $\rho_{\lambda}$ is the 
contribution from the tension of the brane which is simply a Nambu-Goto
term.  From the Israel equations we may then derive the Friedmann equation 
on the brane:
\begin{eqnarray}
\label{Friedmann} 
H^{2}(\tau)& = &{\Lambda_{eff}\over3} - \frac{k}{a^2} +
\left({8\pi G_4\over 3}\right) \rho  \nonumber \\
           &   & +\mbox{} \left({\pi}^2 G_5\right)^2\rho^2+{{2m(\tau,a)} \over
a^4} - \frac{q^2}{a^6}\,,
\end{eqnarray}

\noindent Henceforth, we shall set $q = 0$ and 
$k$ the (spatial) curvature of the brane to unity and
$G_5$ to $ \sqrt{4G_4/3\pi\rho_{\lambda}}$.  
$\Lambda_{eff}$ is
the 4-dimensional cosmological constant on the brane, which is given in terms
of the brane tension $\rho_{\lambda}$ and the bulk cosmological
constant $\Lambda$: $\Lambda_{eff} \equiv \Lambda_4 = \left( \frac{\Lambda}{2} + 4\pi G_4
\rho_{\lambda} \right)$.  

Thus, we find that a time-dependent mass in the bulk gives
rise to a time-dependent term that scales like radiation.  In other words,
the collapse of radiation (to form a black hole) in the bulk gives rise to a
component of `Hot Dark Matter' on the brane.  While the form of (10) is
what one might naively expect given the adS/CFT duality \cite{gubser}, our calculation
actually shows explicitly how a time-dependent mass term in the bulk will
affect the brane-world.

\section{Thermodynamics of the system}

We have seen that the thermal nucleation of a black hole in the bulk 
gives rise to a time-dependent radiation term on the brane.  
To understand how this term will affect brane-world cosmology,
we would like to solve for the back-reaction
generated by 1-loop effects in the bulk.  As this is a rather difficult
problem, we will simply employ the technology of Euclidean quantum gravity.

We want to argue that there is no `runaway' process, whereby a black
hole can keep absorbing more energy than it emits.  Furthermore, we are only
interested in the situation where the brane-world remains outside of the horizon;
from the analysis of \cite{dddw} it is easy to see that this is true only when
$r^4 > 4m^2$.  In other words, the brane-world has to be sufficiently large
relative to the black hole to accomodate the black hole.

The Euclidean action is given as
\begin{equation}
I = -{1\over {{\kappa}_5}^2} 
\int_{\cal M} d^5x\, \sqrt{g}(R - 2{\Lambda})
- {1\over 8\pi} \int_{\partial \cal M} d^4x\,\sqrt{\gamma}K
\end{equation}
where $\cal M$ is the bulk region of the spacetime, with boundary
$\partial \cal M$ (i.e., the boundary is where the brane-world is located).
In the usual situation (where the brane is actually the boundary of AdS)
the boundary term does not contribute; however, in this situation the boundary
is at a finite distance and does contribute.
In order to show that the system equilibrates, we first calculate the action
of the Euclidean section of the brane-world with a black hole in the bulk
(denoted $I_{BH}$), then we calculate the action for the Euclidean section
of the brane-world with no black hole (denoted $I_{BW}$), and we form the
difference:
\[
{\Delta}I = I_{BH} - I_{BW}
\]
Clearly, a phase transition occurs when ${\Delta}I$ changes sign (i.e., 
when the system with a black hole has less action), and likewise the system
is thermally stable when the specific heat, $H$, is positive:
\[
H = -{\beta}{\partial}_{\beta}I
\]
where ${\beta} = T^{-1}$ is the inverse temperature, or period of the
Euclidean section of the relevant solution.  The only subtle point in
performing this calculation lies in matching the two solutions in the 
asymptotic region, near the brane-world.  Once this is done, it is
straightforward to see that the modification to the ordinary Hawking-Page
analysis is negligible, as long as the brane-world remains far from
the horizon (in particular, as long as $r^4 > 4m^2$).  More precisely,
the two boundary contributions differ by $\approx \frac{\sqrt{2m}}{r}$.
Furthermore, even when this correction is non-negligible, it does not affect the phase
structure found by Hawking and Page.

From the point of view of the AdS/CFT duality, this makes sense.
Truncating $AdS_5$ with a de Sitter brane introduces a UV cutoff
in the CFT defined on the brane, but in our approximation it does not
affect the boundary conditions on the bulk radiation (i.e., we assumed
the same boundary conditions as in the original Hawking-Page paper \cite{hawkpage}). 
As long as sufficiently large bulk black holes are allowed, we already knew
that these holes can be in thermal equilibrium with thermal radiation.
The main shortcoming of this analysis is that we assumed reflecting boundary conditions 
for all bulk modes at the brane-world.  Presumably, 
there may exist fields on the brane which can be excited 
by bulk gravitons (for example).  This would mean that the boundary conditions
at the brane would have to allow for some absorption.  We will consider such
boundary conditions in an upcoming paper \cite{tocome}.

To summarize:  We have shown that the Friedmann equations of the second
Randall-Sundrum model will generically contain a hot dark radiation term
which has an intensity that is variable but will 
settle down to a fixed value during inflation.  Once inflation ends,
the mass function will presumably change in some way; it remains to
be shown that the system will still equilibrate once the equation of 
state on the brane is no longer pure vacuum.
It is amusing to note that this varying mass function
is reminiscent of the `$C$-field' introduced by Hoyle and Narlikar in 
their steady state model and later on in the quasi-steady state model of 
the universe (see e.g.~\cite{qssc} and references therein).  Both our 
dynamical mass function and their $C$-field can pump energy into the universe, 
so that total energy on the brane is not conserved.  
We will have more to say about the cosmology of this model in an 
upcoming paper \cite{tocome}.

{\noindent \bf Acknowledgements}\\

We thank R. Battye, E. Bertschinger,  A. Guth, P. Khorsand,
P. Kraus, P. Mannheim, L. Randall and C. Vafa for useful conversations.  
AC and AK are supported in part 
by funds provided by the U.S. Department of Energy (D.O.E.) under
cooperative research agreement DE-FC02-94ER40818.  A.N. is supported by NSF
grant ACI-9619019.  CTP preprint \# 2998.

\end{multicols}

\end{document}